\documentstyle[12pt,epsf]{article}
\newcommand{\lsim}{\,{\buildrel < \over {_\sim}}\,}
\newcommand{\gsim}{\,{\buildrel > \over {_\sim}}\,}

\begin{document}

\begin{titlepage}
\begin{flushright}
JYFL-8/98\\
US-FT/14-98\\
hep-ph/9807297\\
July 1998
\end{flushright}
\begin{centering}
\vfill

{\bf THE SCALE DEPENDENT NUCLEAR EFFECTS IN PARTON DISTRIBUTIONS FOR 
PRACTICAL APPLICATIONS}\\

\vspace{0.5cm}
K.J. Eskola$^{\rm a}$\footnote{kari.eskola@phys.jyu.fi},
V.J. Kolhinen$^{\rm a}$\footnote{vesa.kolhinen@phys.jyu.fi} and
C.A. Salgado$^{\rm b}$\footnote{salgado@fpaxp1.usc.es}

\vspace{1cm}
{\em $^{\rm a}$ Department of Physics, University of Jyv\"askyl\"a,\\
P.O.Box 35, FIN-40351 Jyv\"askyl\"a, Finland\\}

\vspace{0.3cm}

{\em $^{\rm b}$ Departamento de F\'\i sica de Part\'\i culas,\\
Universidade de Santiago de Compostela,\\
E-15706 Santiago de Compostela, Spain.\\}

\vspace{1cm} 
{\bf Abstract} \\ 

The scale dependence of the ratios of parton distributions in a proton
of a nucleus $A$ and in the free proton,
$R_i^A(x,Q^2)=f_{i/A}(x,Q^2)/f_i(x,Q^2)$, is studied within the
framework of the lowest order leading-twist DGLAP evolution. By
evolving the initial nuclear distributions obtained with the GRV-LO
and CTEQ4L sets at a scale $Q_0^2$, we show that the ratios
$R_i^A(x,Q^2)$ are only  moderately sensitive to the choice of a
specific modern set of free parton distributions. We propose that to a
good first approximation, this parton distribution set-dependence of
the nuclear ratios $R_i^A(x,Q^2)$ can be neglected in practical
applications. With this result, we offer a numerical parametrization
of $R_i^A(x,Q^2)$ for all parton flavours $i$ in any $A>2$, and
at any $10^{-6}\le x \le 1$ and any $Q^2\ge 2.25$ GeV$^2$ for computing cross
sections of hard processes in nuclear collisions.
\end{centering}

\vspace{0.3cm}\noindent

\vfill
\end{titlepage}

\section{Introduction}

The measurements of the nuclear structure function $F_2^A(x,Q^2)$ in
deeply inelastic lepton-nucleus scattering (DIS)
\cite{EARLY}-\cite{ARNEODO94} indicate clearly that
parton distributions of the bound protons are different from those of
the free protons, $f_{i/A}(x,Q^2) \neq f_{i/p}(x,Q^2)$.  The nuclear
effects are often categorized according to those observed in the ratio
of the structure functions of nuclei relative to deuterium, $R_{F_2}^A\equiv
F_2^A/F_2^{\rm D}$: shadowing ($R_{F_2}^A\le 1$) at Bjorken-$x \lsim 0.1$,
anti-shadowing ($R_{F_2}^A\ge 1$) at $0.1 \lsim x \lsim 0.3$, EMC
effect ($R_{F_2}^A\le 1$) at $0.3 \lsim x\lsim 0.7$, and Fermi motion
($R_{F_2}^A\ge 1$) towards $x\rightarrow1$ and beyond.
The recent high-precision measurements by the New Muon Collaboration
(NMC) of the structure function $F_2$ of tin vs. that of carbon, 
$F_2^{\rm Sn}/F_2^{\rm C}$ \cite{NMC96}, have also
revealed a $Q^2$-dependence at small values of $x$.

Theoretically, the origin of the nuclear effects is still under debate
but it is believed that different mechanisms are responsible for the
effects in the different regions of $x$. For a compact introduction
and references to the various theoretical models we refer the reader
to Arneodo's review, Ref. \cite{ARNEODO94}.

In this paper, as a sequel to Ref. \cite{EKR98}, we are going to focus
on studying the perturbative QCD-evolution of the ratios $R_i^A\equiv
f_{i/A}(x,Q^2)/f_{i/p}(x,Q^2)$ within the framework of lowest order
(LO) leading twist (LT) DGLAP evolution \cite{DGLAP}. To our knowledge
the perturbative QCD-evolution of nuclear parton densities has been
studied at least in Refs. \cite{QIU}-\cite{AYALA}.  

Our approach is in practice very similar to the case of the free
proton: once the absolute nuclear parton distributions are given at an
initial scale $Q^2=Q_0^2\gg\Lambda_{\rm QCD}^2$ and at some $x_{\rm min}\le
x\le 1$, the further evolution in $Q^2$ is predicted (at this range of $x$) by
the DGLAP equations. In other words, as in \cite{EKR98}, we assume
that at $Q^2\gsim Q_0^2$ the scale evolution of the nuclear parton
densities is purely perturbative and that the initial conditions at $Q_0^2$
contain nonperturbative input. We determine the initial nuclear
parton distributions iteratively through the QCD evolution by using
experimental data and conservation of momentum and baryon number as
constraints. It should be noted that we do not try to explain the
origin of the nuclear effects but we study their behaviour once they
are there at an initial scale. Furthermore, since we neglect all the
higher twist contributions, like the perturbative GLR-MQ terms
\cite{GLRMQ} at small values of $x$, no nuclear effects are generated
through the evolution but they are all hidden in the initial
distributions at $Q_0^2$.

In Ref. \cite{EKR98}, the DIS data on $F_2^A/F_2^{\rm D}$ and
$F_2^A/F_2^{\rm C}$ \cite{NMCre}-\cite{E665} were used together with
the measurements of the Drell--Yan (DY) cross sections in $pA$
vs. $p{\rm D}$ collisions \cite{E772}, and with the conservation of
baryon number and total momentum to determine the initial nuclear
parton distributions at a scale $Q_0^2=2.25$ GeV$^2$. Especially, it
was shown in \cite{EKR98} that the LT LO DGLAP-evolution can account
very well for the $Q^2$ dependence observed in the experimental ratio
$F_2^{\rm Sn}/F_2^{\rm C}$ \cite{NMC96}.

The measurements of the {\em ratios} $R_{F_2}^A$ in DIS, and also the
ratios of nuclear Drell-Yan (DY) cross sections relative to deuterium,
provide us with information on the nuclear parton distributions only
relative to the nucleons in deuterium. In an approximation where
the small nuclear effects in deuterium are neglected, as done in
\cite{EKR98} and as we will do here as well, we get information {\em
relative to the free nucleons}. Then, quite obviously the extraction
of the {\em absolute} nuclear parton distributions depend on our
choice for the set of parton distributions in the free proton, and the
QCD-evolution itself will depend on this choice, too\footnote{We will
refer to the dependence on the choice for the set of parton
distributions in the free proton as ``set-dependence''}. In Ref.
\cite{EKR98}, it was anticipated that the {\em ratios} $R_i^A(x,Q^2)$
should not depend as strongly on the choice for the set as the
absolute distributions do. In this paper, we will verify this
statement quantitatively by comparing the scale-evolved nuclear
effects initially obtained by using the GRV-LO \cite{GRVLO} and CTEQ4L
\cite{CTEQ4L} distributions.

After demonstrating that the set-dependence in the nuclear ratios
$R_i^A(x,Q^2)$ for individual flavours indeed is a negligible effect
as compared to the current overall uncertainties, it becomes more
meaningful to present a numerical parametrization of $R_i^A(x,Q^2)$
for practical applications.  The parametrization can then be used with
any modern set of LO parton distributions to obtain the absolute
nuclear parton distributions needed for computing hard scattering
cross sections in nuclear collisions. The parametrization we offer can be
called for any parton flavour $i$, any $A>2$, $10^{-6}\le x \le 1$ and
$Q^2\ge 2.25$ GeV$^2$. The parametrization, which we give in form of a
short Fortran code, is now available from us\footnote{via email or in
the WWW, http://fpaxp1.usc.es/phenom/}.

\section{The framework and the assumptions}

A detailed formulation of our approach can be found in
Ref. \cite{EKR98} but let us recall the underlying basic assumptions
also here.  The starting point is the measured ratio of structure
functions of a nucleus $A$ and deuterium D, $R_{F_2}^A\equiv
F_2^A/F_2^{\rm D}$. In the one-photon approximation the ratio of the
cross sections measured in $lA$ and in $l{\rm D}$ gives directly the
ratio $R_{F_2}^A$, provided that the ratio of the cross sections of
longitudinally and transversally polarized virtual photons with the
target, $R\equiv\sigma_L/\sigma_T$, does not depend on the target. So
far, experiments have not shown any significant target dependence
\cite{ARNEODO94,NMCsys,NMC96}. Our approach, being currently only in the
lowest order in the cross sections, will not generate such a
dependence, either.

First, the structure function ratio $R_{F_2}^A\equiv F_2^A/F_2^{\rm
D}$ is expressed in terms of nuclear parton distributions by using the
LO relation in the QCD improved parton model, $F_2=\sum_q  e_q^2
[xf_q(x,Q^2)+xf_{\bar q}(x,Q^2)]$.  By the nuclear distributions
$f_{i/A}$ of a parton flavour $i$ we mean the average distributions of
a parton flavour $i$ in a proton of a nucleus $A$:
$f_{i/p/A}(x,Q^2)\equiv f_{i/A}(x,Q^2)$. In principle, the nuclear
parton distributions are non-zero up to $x\rightarrow A$ but the small
tails are completely negligible from the point of view of our
study. We therefore approximate $f_{i/A}(x\ge 1,Q^2)=0$.

For isoscalar nuclei $d_{n/A}= u_{p/A}\equiv u_A$ and
$u_{n/A}=d_{p/A}\equiv d_A$ obviously hold but we have to assume that
these are good approximations for non-isoscalar nuclei as
well\footnote{For non-isoscalar nuclei the corresponding exact
relation is $d(u)_{n/A_Z}= u(d)_{p/A_{A-Z}}$}.  We also neglect the
small nuclear effects in deuterium \cite{BK}. With these
approximations $R_{F_2}^A$ reflects directly the deviations of $F_2^A$
from the free nucleons.

The nuclear effects in the parton distributions are defined through
the ratios for each flavour as,
\begin{equation}
R_i^A(x,Q^2) \equiv \frac{f_{i/A}(x,Q^2)}{f_i(x,Q^2)}
\end{equation}
where $f_i$ is the distribution of the flavour $i$ in the free proton
as given by the chosen set of parton distributions. The modern sets
include the rapid increase of the gluon and sea quark distributions at
small values of $x$, in accordance with the rise discovered in the
structure function $F_2^{ep}$ at HERA \cite{HERA}. In the DGLAP
evolution we treat the massive quarks as massless and generate them
only radiatively above fixed threshold scales.  The two modern LO sets
of parton distributions in which the heavy quarks are treated in this
manner are the GRV-LO \cite{GRVLO} and CTEQ4L \cite{CTEQ4L}, which is
why we use only these two sets in this work.

To eliminate the complication of determining nuclear effects for the
heavy quark distributions at the initial scale $Q_0^2$, we choose
$Q_0^2$ at or below the charm-threshold $Q_c^2$, where $xc=xc_A=0$.
In the set GRV-LO the threshold is $Q_c^2 = 2.25$ GeV$^2$ and in the
CTEQ4L $Q_c^2 = 2.56$ GeV$^2$. To be consistent, we use these in the
corresponding evolution in the nuclear case as well.  The HERA results
have also shown that the leading twist evolution describes the
structure function $F_2^{ep}$ well at $Q^2\gsim 1$~GeV$^2$ and $x\gsim
10^{-4}$ \cite{HERAlowQ}.  In the nucleus, however, the higher twist
effects can be expected to be stronger \cite{GLRMQ,QIU,EQW}. Since we
will not include these here, we should not start the
perturbative evolution of nuclear parton distributions below $Q^2\lsim
1$ GeV$^2$. Therefore, for not too small values of $x$, it is fairly
safe to choose $Q_0^2=Q^2_c=2.25$ GeV$^2$ of the set GRV-LO.  The
results in Ref.  \cite{EKR98} show that this is a valid initial scale
at least for $x\gsim 0.01$, where the higher twist contributions are
negligible.

As further approximations, but for the initial conditions at $Q^2=Q_0^2$
only, it is assumed that the differencies between the nuclear effects
for sea quarks and antiquarks, as well as for different sea quark
flavours are negligible: $R_{\bar q}^A(x,Q_0^2)=R_{q_{\rm
sea}}^A(x,Q_0^2)=R_S^A(x,Q_0^2)$.  Similarly, the same nuclear effects 
are assumed for the valence quarks at the initial scale,
$R_{u_V}^A(x,Q_0^2)=R_{d_V}^A(x,Q_0^2)=R_V^A(x,Q_0^2)$.  The fact that with
these approximations we get a stable evolution -- i.e. that the nuclear
effects do not rapidly evolve away from what is assumed at $Q_0^2$ --
shows that these are indeed  reasonable first approximations.

Then, we arrive at a simple formula for an isoscalar nucleus,
\begin{equation}
R_{F_2}^A(x,Q^2_0)= A_V^{IS}(x,Q^2_0) R_V^A(x,Q^2_0) +
[A_{ud}^{IS}(x,Q^2_0) + A_s(x,Q^2_0)]  R_S^A(x,Q^2_0),
\label{RF2appro}
\end{equation}
where the coefficients are known exactly \cite{EKR98}:
\begin{eqnarray}
A_V^{IS}(x,Q^2) &=& 5[u_V(x,Q^2)+d_V(x,Q^2)]/N_{F_2}(x,Q^2)\\
A_{ud}^{IS}(x,Q^2) &=& 10[\bar u(x,Q^2)+\bar
d(x,Q^2)]/N_{F_2}(x,Q^2)\\
A_s(x,Q^2) &=& 4s(x,Q^2)/N_{F_2}(x,Q^2)\\
N_{F_2}(x,Q^2)&=& 5[u_V(x,Q^2) + d_V(x,Q^2)] +10[\bar u(x,Q^2)+\bar d(x,Q^2)]
             + 4s(x,Q^2).
\end{eqnarray}
They obviously depend on the specific set of parton distributions
chosen for the free proton.  The above Eq.~(\ref{RF2appro}) correlates
the nuclear effects of the valence quarks with those of the sea quarks
through the ratio $R_{F_2}^A$. As described in detail in
\cite{EKR98}, we fix $R_{F_2}^A$ iteratively through the evolution by
using the DIS data \cite{NMCre}-\cite{E665}, decompose $R_{F_2}^A$
into $R_V^A$ and $R_S^A$ at $Q^2=Q_0^2$ according to
Eq.~(\ref{RF2appro}) by simultaneously constraining $R_V^A$ by
conservation of baryon number and $R_S^A$ by the DY data \cite{E772}.
At large values of $x$ the sea quark distributions are very small as
compared to the valence distributions, and the contribution from the
sea to $R_{F_2}^A$ is negligible. This is why we have no constraints for
$R_S^A$ at $x\gsim0.3$. We simply assume that the sea quarks show a
similar EMC effect as the valence quarks do, so at large values of $x$ we
have $R_V^A=R_S^A=R_{F_2}^A$.

Also the region below $x\lsim 0.01$ is not well constrained: there
exist experimental DIS data, even at much lower values of $x$
\cite{NMCsat,E665sat}, but due to the correlation of $x$ and $\langle
Q^2\rangle$ in the experiments, the values of $\langle Q^2\rangle$ are
in the nonperturbative region from our point of view.  Also, as pointed out in
\cite{EKR98}, since $\partial R_{F_2}^A(x,Q^2)/\partial \log Q^2 \sim
R_G^A(2x,Q^2)-R_{F_2}^A(x,Q^2)$, the slope of $R_{F_2}^A$ in $Q^2$ --
even the initial sign of the slope -- depends on what is assumed for
the initial nuclear gluon profile $R_G^A(x,Q_0^2)$ at small values of
$x$.  We assume a saturation of shadowing in $R_{F_2}^A$, as observed
in the nonperturbative region \cite{NMCsat,E665sat}, and again, to
reach a stable evolution \cite{EKR98}, take $R_G^A=R_{F_2}^A$ at small
values of $x$.  In relation to the data on $R_{F_2}^A$ measured in the
region of very small $x$ \cite{NMCsat,E665sat}, we basically expect
that the nonperturbative evolution does not change the sign of the 
$Q^2$-slope of $R_{F_2}^A$. This interesting question is related 
to the origin of nuclear shadowing. Within our approach, however, we
cannot address this problem further.

Apart from the overall momentum conservation, we do not,
unfortunately, have any direct constraints for the nuclear gluon
distributions.  In principle, as pointed out in \cite{KJE}, the $Q^2$
dependence of $R_{F_2}^A$, if measured very accurately at small enough
values of $x$, would serve as such a constraint.  As a step towards
this direction, the NMC has recently measured the $Q^2$ dependence of
$F_2^{\rm Sn}/F_2^{\rm C}$ \cite{NMC96} and Pirner and Gousset \cite{PIRNER}
have used this data \cite{MUCKLICH} to extract the approximate nuclear
gluon densities.  Since we now assume a similar shadowing in $R_G^A$
as in $R_{F_2}^A$, we are bound to have some antishadowing due to the
overall momentum conservation \cite{EKR98}. In our analysis, we use
the results of Ref. \cite{PIRNER}, to constrain only the point where
$R_G^A(x,Q_0^2)\approx 1$. By requiring again a stable evolution
(unlike in \cite{KJE}) we expect that the gluons have an EMC-effect as
well.  The height of the antishadowing peak in $R_G^A(x,Q_0^2)$ is
then fixed by the momentum conservation. As shown in \cite{EKR98},
with the full scale evolution this results in quite a good overall
agreement with the analysis in Ref. \cite{PIRNER}.

\section{The set-dependence of the nuclear effects}

In this section, we will make a quantitative comparison of the
set-dependence of the nuclear effects for all parton flavours.  More
exactly, we will compare the results of \cite{EKR98} obtained with the
parton distribution set GRV-LO \cite{GRVLO} to those obtained with the
set CTEQ4L\cite{CTEQ4L}.  

First, the set-dependence will obviously be the largest for largest
nuclei, so it is sufficient to focus on the set-dependence for an
isoscalar $A=208$.  To have a consistent but the most straightforward
comparison, we fix the parametrization of $R_{F_2}^A(x,Q_0^2)$ to be
exactly the same as given in \cite{EKR98}. By doing this, we
anticipate that the set-dependence of the scale evolution of
$R_{F_2}^A$ will be negligible, so that a new iteration with a new
initial parametrization of $R_{F_2}^A(x,Q_0^2)$ is not
needed. Similarly, we keep the parameters for $R_V^A(x,Q_0^2)$ at
$x\lsim 0.1$ the same as given in \cite{EKR98}. 

\begin{figure}[tb]
\vspace{-1.5cm}
\centerline{\epsfxsize=15cm\epsfbox{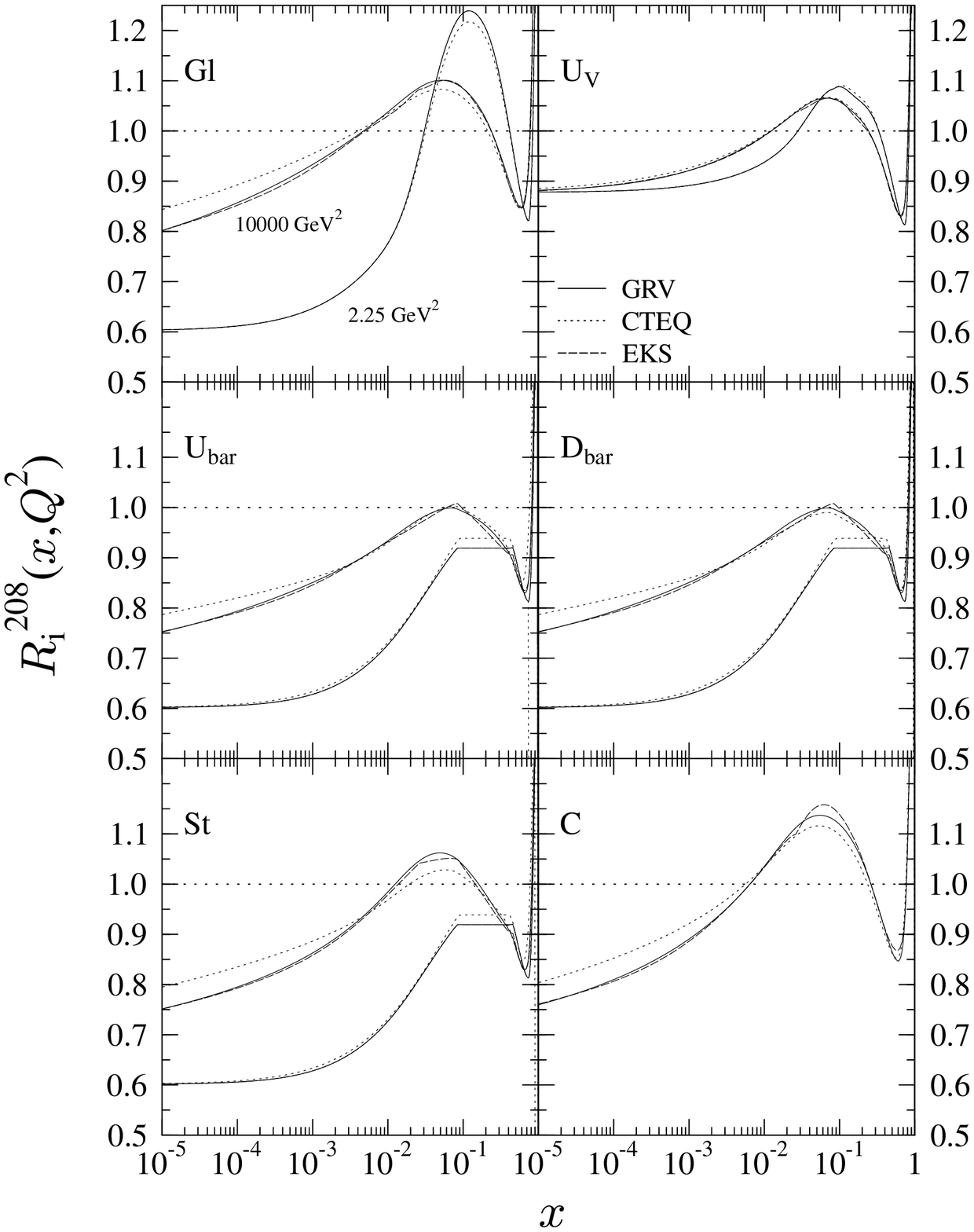}}
\vspace{-2cm}
\caption[a] { {\small The nuclear ratios $R_i^A(x,Q^2)$ for individual
parton flavours $i=g, u_V, \bar u, \bar d, s, c$ of a lead
nucleus $A=208$ as functions of $x$ at fixed values of
$Q^2=Q_0^2=2.25$ GeV$^2$ and $Q^2=10000$ GeV$^2$ as obtained by using
the GRV-LO \cite{GRVLO} distributions (solid lines) and the CTEQ4L
\cite{CTEQ4L} distributions (dotted lines) for the free proton.  The
dashed lines show our numerical parametrization (EKS) to the nuclear
effects obtained in the GRV-LO case. The ratios $R_{d_V}^A$ are almost
identical to $R_{u_V}^A$, and are not shown.  The charm ratios are
presented only for $Q^2=10000$ GeV$^2$, since the charm distributions
are generated only above our $Q_0^2$. The ratios $R_b^A$ at
$Q^2=10000$ GeV$^2$ behave as $R_c^A$, and are not shown.  For
practical purposes the set-dependence of the ratios $R_i^A(x,Q^2)$ is
negligible.  }}
\label{SETDEP}
\end{figure}

To apply the initial parametrization of $R_{F_2}^A(x,Q_0^2)$ to the
CTEQ4L distributions in a consistent manner, and to remove a source of
uncertainty, we first evolve the CTEQ4L distributions {\em downwards}
from $Q^2=Q_0^{2,\rm CTEQ4L}=2.56$ GeV$^2$ to our starting scale
$Q^2=Q_0^2=2.25$ GeV$^2$, and only then apply the procedure of
extracting the initial nuclear effects in the valence quarks,
$R_V^A(x,Q_0^2)$, in the sea quarks, $R_S^A(x,Q_0^2)$, and in the
glue, $R_G^A(x,Q_0^2)$.  These initial ratios can be found in
Fig.~\ref{SETDEP}, together with the previous results \cite{EKR98}
with the set GRV-LO. The resulting set-dependence in $R_S^A(x,Q_0^2)$
and $R_V^A(x,Q_0^2)$ is small and certainly well within the expected
overall uncertainty in fixing the initial conditions for these ratios.

Notice also that the set-dependence of the initial gluon ratio
$R_G^A(x,Q_0^2)$ is conveniently small, as seen in Fig.~\ref{SETDEP},
even though the {\em absolute} gluon distributions of GRV-LO and
CTEQ4L differ from each other, even by a factor $\sim$2. Remember that the
shadowing of $R_G^A(x,Q_0^2)$ at very small values of $x$ was fixed to
coincide with $R_{F_2}^A(x,Q_0)$ but the antishadowing peak in
$R_G^A(x,Q_0^2)$ is determined from the overall momentum conservation.

Next, we perform the DGLAP evolution \cite{DGLAP} for the absolute
nuclear parton distributions now obtained by using the set CTEQ4L, and
form the nuclear ratios of individual ratios at a large scale
$Q^2=10000$ GeV$^2$. The comparison with the ratios obtained by the
set GRV-LO is shown in Fig.~\ref{SETDEP}. It is quite interesting to
notice that even if the absolute distributions may differ from each
other considerably, in the scale-evolved nuclear ratios the
differencies tend to cancel out, making the set-dependence a small
effect. The set-dependence is the largest for the gluon ratios at
small values of $x$, and this is transmitted directly to the sea
quarks and to the heavy quarks. The strange-quarks also show a
set-dependence of at most 5 \%. Another interesting point is that the
difference between $R_{\bar u}^A$ and $R_{\bar d}^A$ stays quite small
even though $\bar u\ne\bar d$ for the set CTEQ4L.

Bearing in mind the uncertainties in fixing the initial ratios,
especially the assumptions regarding shadowing at $x\lsim 0.01$, and
the gluon profile in general, the conclusion is that the
set-dependence is a negligible effect as compared to the current
overall uncertainty in determining the initial conditions.

\begin{figure}[tb]
\vspace{-1.5cm}
\centerline{\epsfxsize=16cm\epsfbox{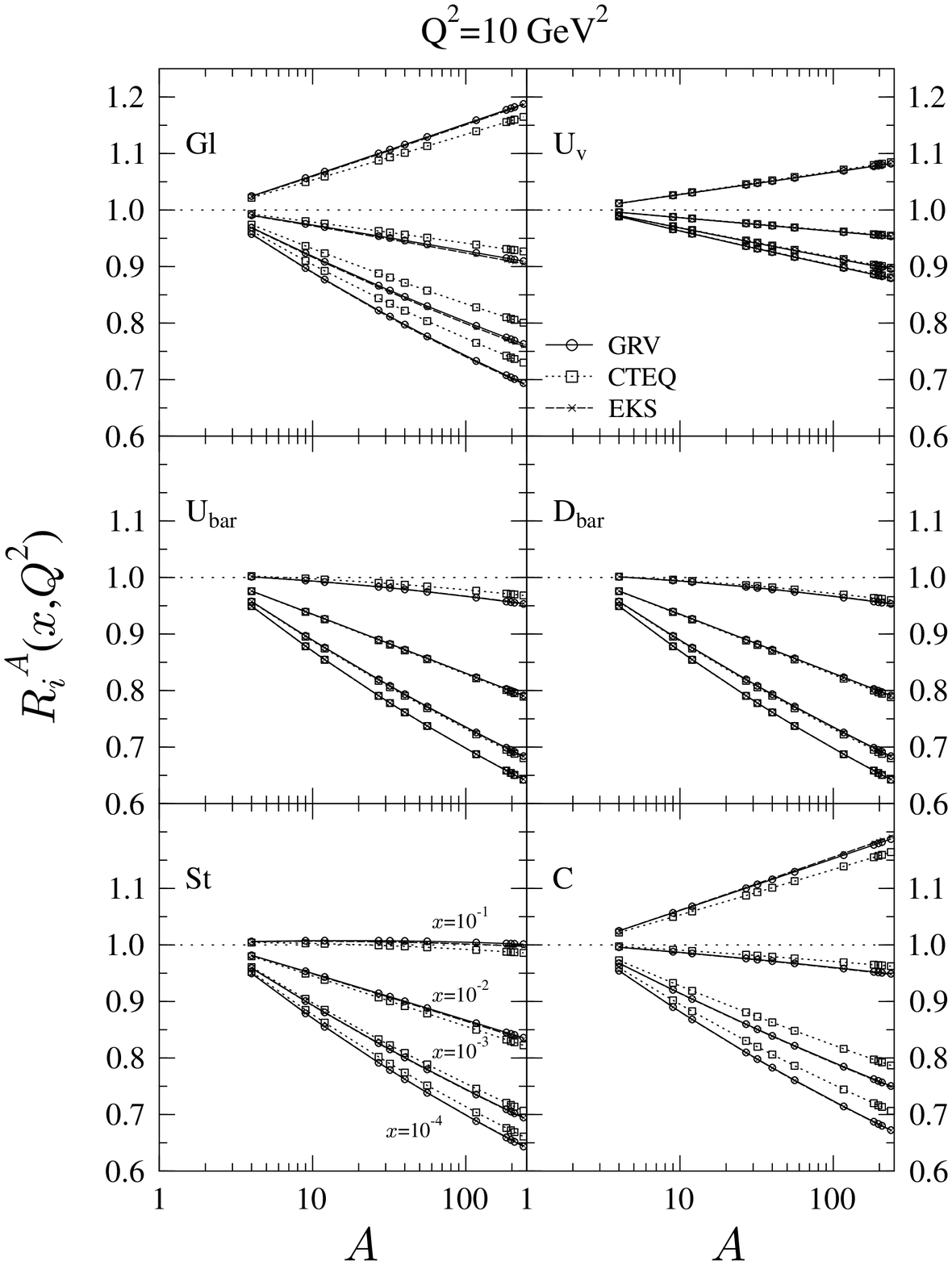}}
\vspace{-1cm}
\caption[a] { {\small The nuclear ratios $R_i^A(x,Q^2)$ for individual
parton flavours $i=g, u_V, \bar u, \bar d, s, c$ as
functions of the mass number $A$ at a fixed value of $Q^2 = 10$
GeV$^2$ and at fixed values of $x= 10^{-4},\, 10^{-3},\, 10^{-2},\,
10^{-1}$ (see the panel for the strange quarks). The notation of the
curves is the same as in Fig.~\ref{SETDEP} and the markers show the
nuclei for which we have made the computation.  The set-dependence is
the largest for large nuclei but keeps within $\sim$5 \%.  }}
\label{ADEP}
\end{figure}

In Fig. ~\ref{ADEP}, the set-dependence of the nuclear ratios is shown
as function of nuclear mass $A$ at different fixed values of $x$ and
one fixed value of $Q^2$. This figure demonstrates the fact that for
smaller nuclei the set-dependence is even a smaller effect.

\section{The parametrization for practical applications}

Originally \cite{KJE,EKR98}, we have stored the $x$- and
$Q^2$-dependent nuclear ratios, $R_i^A(x,Q^2)$ for all relevant parton
flavours in a big table for each nucleus separately. In practical
applications of computing cross sections of hard scatterings in
nuclear collisions, these big tables have to be interpolated to obtain
the absolute nuclear distributions\footnote{Also these tables with an
interpolation routine is available from us via email.}
$f_{i/A}(x,Q^2)=R_i^A(x,Q^2)f_i(x,Q^2)$.  Then, in principle, one
should store $\sim 200$  tables to be able to get nuclear parton
distributions in an arbitrary nucleus $A\ge 2$. To avoid this
complication, and to make the nuclear effects more easily available
for any user, we have prepared a numerical Fortran code which
parametrizes the $x$-, $Q^2$- and especially the $A$-dependence of
$R_i^A(x,Q^2)$ for all parton flavours $i$ of an
isoscalar nucleus $A$. 

We offer the numerical parametrization of the ratios $R_i^A(x,Q^2)$ in
two formats: The first one, called ``EKS98'', computes the nuclear
ratios for all parton flavours within one function call, and the second
one, called``EKS98r'', computes the nuclear ratio of the parton flavour
specified by the user. Both versions are initialized by reading in
two small tables. The codes and the tables are contained in a package 
now available from any of the authors via email. For an easier access,
the package is also placed in the WWW\footnote{http://fpaxp1.usc.es/phenom/}.

Here we do not wish to go into the numerical details of making such a
parametrization. Instead, we simply demonstrate the quality of our
parametrization in Figs.~(\ref{SETDEP}) and (\ref{ADEP}). As can be seen
in the figures, the parametrization is based on the results
\cite{EKR98} obtained with the set GRV-LO. Therefore the
parametrization is closer to those results and the region of validity
is $10^{-6}\le x\le 1$ and $Q^2\ge 2.25$ GeV$^2$. For $A\le2$, the
parametrization simply returns unity, in accordance with the fact that
nuclear effects in deuterium were neglected.

Now, with the results of the previous section on the approximate
set-independence of the ratios $R_i^A$, the absolute parton
distributions in protons of an arbitrary nucleus $A$ with $Z$ protons
and $A-Z$ neutrons can be obtained simply by multiplying the parton
distributions $f_i^{\rm PDset}$ of any modern lowest order set by our
parametrization $R_i^{A,{\rm EKS98}}$,
\begin{equation}
f_{i/A}^{\rm PDset}(x,Q^2)\equiv f_{i/p/A}^{\rm PDset}(x,Q^2) 
\approx R_i^{A,{\rm EKS98}}(x,Q^2) f_i^{\rm PDset}(x,Q^2),
\end{equation}
and the corresponding distributions in a bound neutron can be obtained
through the assumed isospin symmetry (exact for isoscalars, though).
Thus, our parametrization should be easy to use together with parton
distributions from e.g. the CERN PDFLIB \cite{PDFLIB} or other
similar packages.

\section{The set-dependence in the physical quantities}

As the last task, we check explicitly the set-dependence in the
physical quantities studied in Ref. \cite{EKR98}, and simultaneously
we show the range of uncertainty when using our parametrization with
different sets of parton distributions in the free proton instead of
the nuclear ratios obtained specifically for the chosen
set. For brevity, let us call the latter ones as GRV-ratios and
CTEQ-ratios.

\begin{figure}[tb]
\vspace{-1cm}
\centerline{\hspace*{0cm} \epsfxsize=15cm\epsfbox{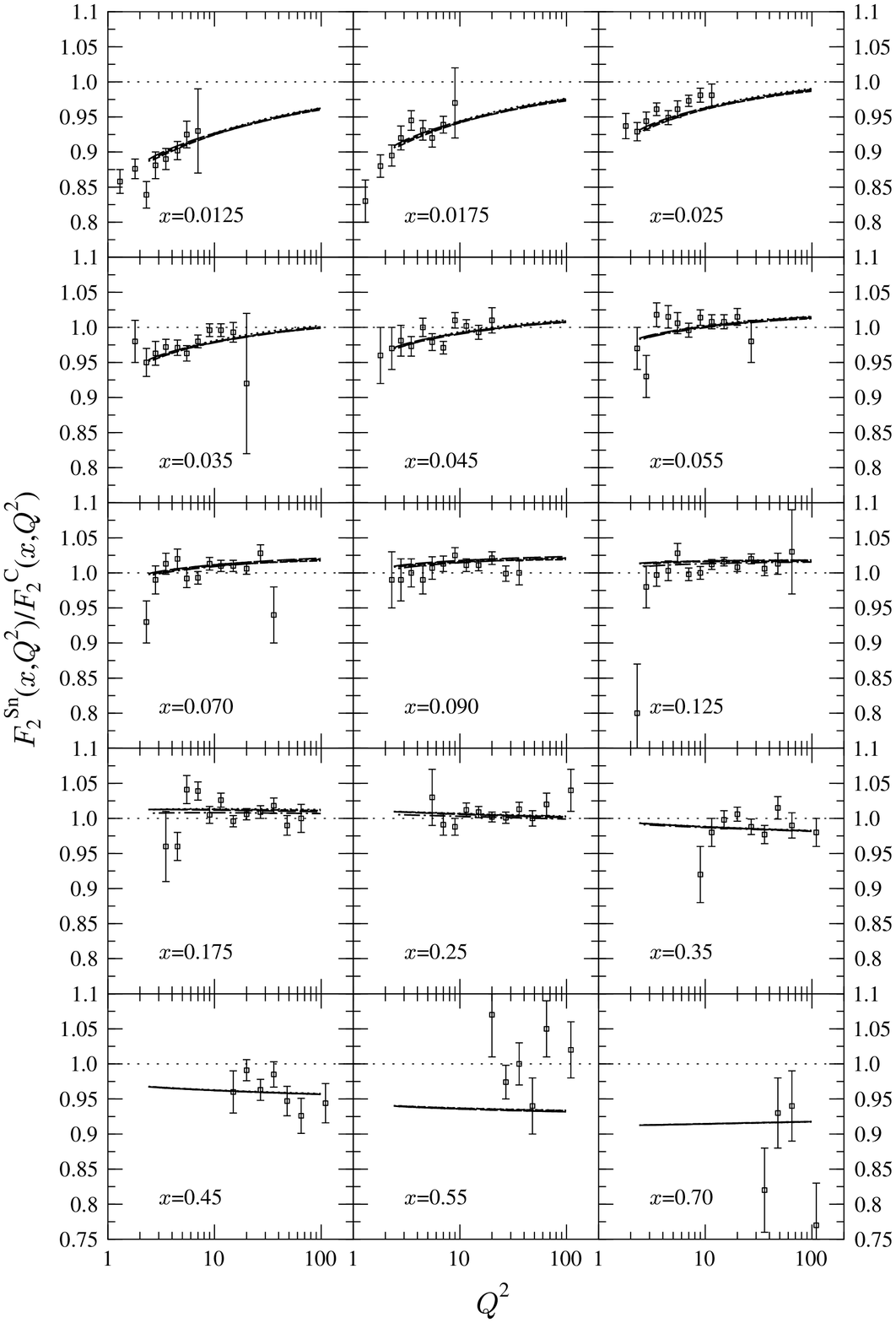}}
\vspace{-0.5cm}
\caption[a] { {\small The ratio $F_2^{\rm Sn}/F_2^{\rm C}$ as a function
of $Q^2$ at several different fixed values of $x$. The data is from
\cite{NMC96}.  The figure contains four calculated curves: two of the
curves correspond to the ``exact'' results obtained with the nuclear
ratios for the GRV-LO distributions (as in \cite{EKR98}) and for the
CTEQ4L-distributions separately, and two curves correspond to the
results obtained with the GRV-LO and CTEQ4L distributions multiplied
by our numerical parametrization of $R_i^A(x,Q^2)$.  There is no
significant difference between the calculated curves.  }}
\label{QDEP}
\end{figure}

First, in Fig.~\ref{QDEP}, we calculate the $Q^2$ dependence of the
ratio $F_2^A/F_2^{\rm C}=R_{F_2}^A/R_{F_2}^{\rm C}$ for an isoscalar
$A=118$, corresponding to an isoscalar tin, and compare our calculation
with the NMC data \cite{NMC96}. There are four curves in
the figure: two curves which have been obtained by using the nuclear
parton distributions obtained with the GRV-ratios and
CTEQ-ratios,  and two curves obtained with our numerical
parametrization.  The four curves practically lie on top of each
other, so the uncertainty due to the choice of the set, and the
uncertainty in our parametrization, are negligible as compared to the
experimental (statistical) error bars.  We have checked that we
arrive in the same conclusion also for other ratios
$R_{F_2}^A(x,Q^2)$, as expected due to the small
differencies of the ratios in Fig.~\ref{SETDEP}.

Another quantity we study here is the ratio of the lowest order Drell-Yan
cross section in $pA$ and $p{\rm D}$ and its comparison to the E772-data
\cite{E772}. As presented in \cite{EKR98}, the cross section ratio for
a nucleus $A$ with $Z$ protons and $A-Z$ neutrons can be written as
\begin{eqnarray}
&\!\!\!\!\!R&\!\!\!\!\!_{DY}^A(x_2,Q^2) \equiv  \frac
{\frac{1}{A} {d\sigma^{pA}_{DY}}/{dx_2dQ^2}}
{\frac{1}{2} {d\sigma^{pD}_{DY}}/{dx_2dQ^2}}
\nonumber\\
&&\!\!\!\!\!\!\!
 = \{4[u_1(\bar u_2^A+\bar d_2^A)+ \bar u_1(u_2^A+d_2^A)] +
  [d_1(\bar d_2^A+\bar u_2^A) + \bar d_1(d_2^A+u_2^A)] +
4s_1s_2^A +...\}/N_{DY}  \nonumber\\
&&\!\!\!\!\!\!\!+ (\frac{2Z}{A}-1)
\{4[u_1(\bar u_2^A-\bar d_2^A) + \bar u_1(u_2^A-d_2^A)]+
[d_1(\bar d_2^A-\bar u_2^A) +  \bar d_1(d_2^A-u_2^A)]\}/N_{DY}
\label{DRELLYAN}
\end{eqnarray}
where the denominator is 
\begin{equation}
N_{DY} = 4[u_1(\bar u_2+\bar d_2)+\bar u_1(u_2+d_2)] +
  [d_1(\bar d_2+\bar u_2) +\bar d_1(d_2+u_2)] + 4s_1s_2 + ...
\end{equation}
and where we have used the notation $q_i^{(A)}\equiv q_{(A)}(x_i,Q^2)$ for
$i = 1,2$ and $q=u,d,s,...$. The scale in the parton distributions
 is the invariant mass $Q^2$ of the lepton pair.  The target 
(projectile) momentum fraction is
$x_2\ (x_1)$, and $x_1 = Q^2/(s x_2)$.

\begin{figure}[htb]
\vspace{-0.5cm}
\centerline{\hspace*{0cm} \epsfxsize=17cm\epsfbox{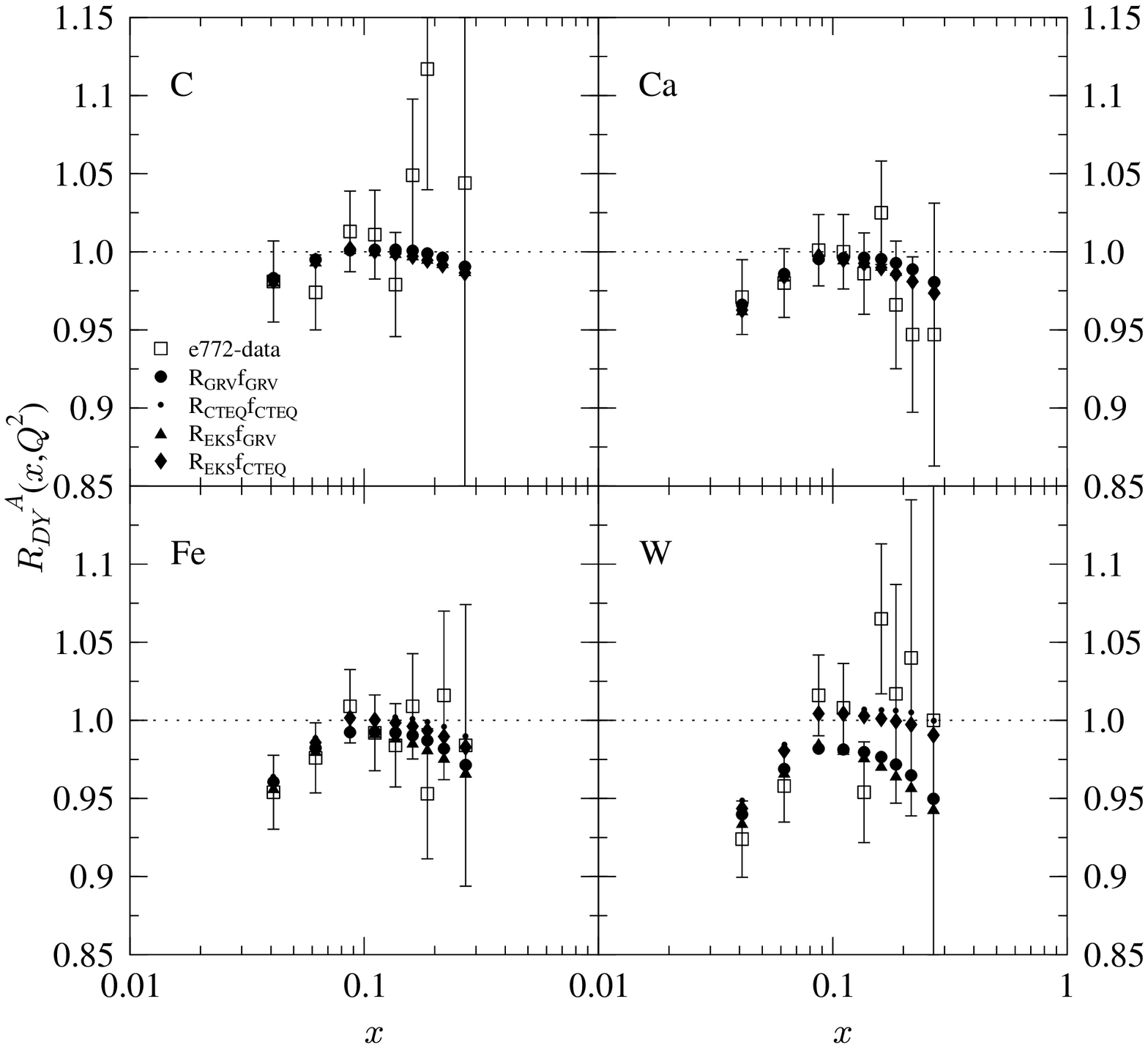}}
\vspace{-5cm}
\caption[a] {{\small The ratio of differential Drell-Yan cross
sections in p$A$ and pD from Eq.~(\ref{DRELLYAN}) as a function of
$x=x_2$ for $^{12}_{~6}$C, $^{40}_{20}$Ca, $^{56}_{26}$Fe and
$^{184}_{~74}$W. The open squares show the E772-data \cite{E772}, and
the filled symbols stand for the calculated ratios $R_{DY}^A(x,Q^2)$ at 
$(x,Q^2)$ corresponding to the experimental values \cite{PMcG}.  The
circles show $R_{DY}^A$ as computed with the nuclear ratios obtained
separately for the GRV-LO set (big circles) and for the CTEQ4L set
(small circles). The results obtained by using our numerical
parametrization (EKS) of $R_i^A$ together with the sets GRV-LO and
CTEQ4L are shown by triangles and diamonds, correspondingly.  As seen
from the panel for tungsten, the differencies between the two parton
distribution sets used for the free proton are larger than the error
from using the set-independent parametrization for the nuclear effects
$R_i^A$.  }}
\label{DY}
\end{figure}

In Fig.~\ref{DY}, we plot the ratio $R_{DY}^A$ for $^{12}_{~6}$C,
$^{40}_{20}$Ca, $^{56}_{26}$Fe and $^{184}_{~74}$W as a function of
the target momentum fraction $x_2$. As in Fig.~\ref{QDEP}, we show
four different calculation in each panel, corresponding to $x_2$ and
$\langle Q^2\rangle$ of the data \cite{PMcG}: two of them are obtained
by using the GRV-ratios and the CTEQ-ratios with the corresponding
sets for the free proton, and the remaining two by using our numerical
parametrization together with the GRV-LO and CTEQ4L parton
distributions for the free proton. Carbon and calcium are isoscalar
nuclei, $A=2Z$, so for them the latter part in Eq.~(\ref{DRELLYAN})
drops out. The remaining small difference in the calculated results
(the big circles and the diamonds) is then the uncertainty in our
parametrization due to the set-dependence.  With iron and especially
with tungsten the situation becomes more interesting because the
non-isoscalar effects start to play a visible role.  In the panel for
tungsten, we clearly see that in the calculated quantities, there are
separate ``error bands'' for the results obtained with GRV-LO and CTEQ4L
distributions. The ``width'' of these bands is again the uncertainty
related to using our parametrization instead of the ``exact'' GRV- or
CTEQ-ratios. As compared to the differencies in the free proton
distributions and to the experimental error bars for $R_{DY}^A$, the
uncertainty due to using our parametrization instead of the ``exact''
nuclear ratios is small.

\section{Discussion and conclusions} 

By using the approach of Ref. \cite{EKR98}, we have studied the
leading twist lowest order DGLAP evolution of nuclear parton
densities. The nuclear effects to the initial parton densities for
free proton at $Q^2=Q_0^2$ have been determined by using the data from
the deeply inelastic $lA$ scatterings \cite{NMCre}-\cite{E665} and
from the Drell-Yan measurements in $pA$ collisions\cite{E772}, and
conservation of baryon number and momentum as constraints. The nuclear
parton distributions, $f_{i/A}^{\rm PDset}(x,Q^2)$, obtained with
different sets may differ considerably in absolute magnitude but in
the scale dependent the nuclear ratios, $R_i^A(x,Q^2)=f_{i/A}^{\rm
PDset}(x,Q^2)/f_i^{\rm PDset}(x,Q^2)$ the differencies between the
sets are much smaller.  By using the GRV-LO \cite{GRVLO} and CTEQ4L
\cite{CTEQ4L} distributions, we have verified that the set-dependence
in $R_i^A(x,Q^2)$ for individual parton flavours is negligible as
compared to the current theoretical overall accuracy in fixing the
initial nuclear effects at $Q^2=Q_0^2$, and also compared to the precision
of the experimental data.

With the result that the dependence of the nuclear ratios on the
choice for the parton distributions for the free proton is negligible,
we have prepared a numerical parametrization of the ratios
$R_i^A(x,Q^2)$ to be used for any practical applications of computing
cross sections of hard scatterings in nuclear collisions, where
absolute nuclear parton distributions, $f_{i/A}^{\rm
PDset}(x,Q^2)=R_i^A(x,Q^2)f_i^{\rm PDset}(x,Q^2)$ are needed.  
We distribute our parametrization in a package containing two alternative
Fortran routines. The package is now available from us via email or 
from the WWW\footnote{http://fpaxp1.usc.es/phenom/}.

In Ref. \cite{EKR98} it was shown that the pure leading twist, lowest
order DGLAP evolution can account very well for the observed scale
dependence of the structure function ratio $F_2^{\rm Sn}/F_2^{\rm C}$.
Here we have demonstrated by using the GRV-LO and CTEQ4L distributions
that using any modern LO set of parton distributions for the free proton
will result in the same conclusion.

We have also shown that quite different isospin effects can be
expected in the ratios of Drell-Yan cross sections in $pA$ vs. $p{\rm D}$
for large non-isoscalar nuclei like tungsten when using sets with
different relative magnitudes for $\bar u$ and $\bar d$.  We have also
demonstrated that it is again sufficient to use set-independent nuclear
ratios, i.e. our parametrization, in the computation of the DY
ratios, the differencies result from the differencies between the parton
distributions of the free proton.

We conclude by recalling the questions in our approach that are still
open regarding the determination of the initial distributions at
$Q^2=Q_0^2$. 

At large values of $x$ the existence of the EMC-effect in the sea and
for the gluon distributions is an assumption.  To directly verify this
experimentally may be quite difficult regarding the gluons. For the
sea quarks, however, there may be a chance to have further constraints
from the NA50 data for Drell-Yan in Pb--Pb collisions at $\sqrt
s=17.2$ GeV at the CERN-SPS, provided that the precision in the
invariant mass distribution will be sufficient at large masses.  If
the sea quarks show an EMC effect, this should serve as an additional
constraint for the gluons through the evolution.

The region at $x\lsim 0.01$ involved assumptions of saturation of
shadowing, and at very small values of $x$ the gluons were expected to
have the same shadowing as $F_2^A$ (or the sea quarks).  To pin down
the nuclear gluon shadowing at small values of $x$, high precision
data on the $Q^2$ dependence of $F_2^A/F_2^{\rm D}$, $F_2^A/F_2^{\rm
C}$ would be needed especially at $x\gsim 0.01$, where the gluons
start to determine the evolution of the sea quarks more strongly
\cite{KJE,EKR98}.  In the future, perhaps also the gluon initiated
processes like open charm and $J/\Psi$ production in DIS
\cite{JPSI} and in p$A$ collisions \cite{E789D,LEITCH} could help
in constraining the gluon antishadowing peak further.

Related to the small-$x$ physics, the role of higher twist effects like
the GLRMQ fusion corrections \cite{GLRMQ} needs still to be studied further 
together with the results from HERA. Also the NLO analysis, both in the cross 
sections and in the evolution, should be performed. Most importantly,
however, our analysis would benefit from an  inclusion of a more 
quantitative estimates of uncertainties and propagation of errors.

\bigskip

\noindent{\bf Acknowledgements.}  We gratefully acknowledge
C. Louren\c{c}o's efforts to get this program started. We also thank
V. Ruuskanen for useful discussions and comments on the manuscript.
We thank N. Armesto and the Hard Probe Collaboration, especially
R. Vogt, for the useful suggestions to improve our numerical
parametrization.  KJE also thanks the Institute of Nuclear Theory,
University of Washington, Seattle, for hospitality during the Hard
Probe workshop in May 1998, where the results of this work were
reported.  CAS is grateful to the PPE Division of CERN (ALICE group)
for hospitality during September - December 1998, when part of this
work was completed.  CAS thanks Xunta de Galicia for financial
support. This work was supported by the Academy of Finland, grant
no. 42376.

\end{document}